\documentstyle[12pt,amssymb,amsfonts]{article}

\def\e{\mbox{e}}

\begin{document}
\begin{flushright}
INR-2000-17\\
UNIL-IPT/00-12
\end{flushright}
\begin{center}
  {\Large\bf Brane world: disappearing massive matter} \\
  \medskip S.~L.~Dubovsky$^a$, V.~A.~Rubakov$^a$, P.~G.~Tinyakov$^{a,b}$\\
    \medskip
  {\small
     $^a$Institute for Nuclear Research of
         the Russian Academy of Sciences,\\  60th October Anniversary
  Prospect, 7a, 117312 Moscow, Russia\\
$^b$Institute of Theoretical Physics,
University of Lausanne, CH-1015 Lausanne, Switzerland\\
  }

\begin{abstract}
In a brane (domain wall) scenario with an infinite extra dimension and
localized gravity, bulk fermions and scalars often have bound states
with zero 4-dimensional mass. In this way massless matter residing on
the brane may be obtained. We consider what happens when one tries to
introduce small, but non-vanishing mass to these matter fields. We
find that the discrete zero modes turn into quasi-localized states with
finite 4-dimensional mass and finite width. The latter is due to
tunneling of massive matter into extra dimension. We argue that this
phenomenon is generic to fields that can have bulk modes. We also
point out that in theories meant to describe massive scalars, the
4-dimensional scalar potential has, in fact, power-law behavior at
large distances.
\end{abstract}
\end{center}

\section{Introduction and summary} 

It has been suggested long time ago \cite{RS,Akama} that localization
of particles on a defect in a higher-dimensional space may serve as an
alternative to the standard Kaluza-Klein compactification.  The
simplest example of such a defect is a domain wall in $(4+1)$
dimensions.  In the domain wall scenario, extra dimension is infinite,
with the observed 4-dimensional fields being zero modes of bulk fields
in the domain wall background.  These zero modes are localized around
the domain wall and thus behave, at low energies, as 4-dimensional
massless fields. Explicit field theoretic realization of the
localization scenario in theories without gravity was straightforward
in the cases of scalars and fermions \cite{RS}; localization of gauge
bosons is much more difficult \cite{DvaliShifman}.

An interesting recent development concerns the gravitational sector
\cite{RSd} (see refs. \cite{str1,str2,str3} for extension to 6
dimensions).  With fine-tuning between (negative) bulk cosmological
constant and (positive) brane tension, there exists a thin-brane
solution to the 5-dimensional Einstein equations which has flat
4-dimensional hypersurfaces,
\begin{equation}
\label{RSmetric}
ds^2=a^2(z)\eta_{\mu\nu}dx^{\mu}dx^{\nu}-dz^2. 
\end{equation}
Here 
\[
a(z) = \exp (-k|z|),
\]
and the parameter $k$ is determined by the 5-dimensional Planck mass
and bulk cosmological constant.  It has been found that the
gravitational field perturbations about the background
(\ref{RSmetric}) have a localized zero mode, a four-dimensional
graviton. Although continuum modes are arbitrarily light in this case,
their interactions with matter on the brane are suppressed. As a
result, gravity experienced by matter residing on the brane is
effectively four-dimensional at distances $r \gg k^{-1}$
\cite{RSd,GarrigaTanaka,GKR}.

It has been shown also \cite{Giga,GKR} that massless bulk scalars in
the Randall--Sundrum background (\ref{RSmetric}) have similar
properties as gravitons: there exists a localized zero mode and a
Kaluza--Klein continuum of arbitrarily light states weakly interacting
with matter residing on the brane. Finally, the original mechanism of
the localization of massless fermions on a domain wall
\cite{RS,JackiwRebbi} works, in a range of parameters, in curved space
(\ref{RSmetric}) as well \cite{Giga}. 

In order to serve as a prototype for any realistic model, the brane
construction has to be supplemented with a mechanism of mass
generation for the 4-dimensional fields. In the usual Kaluza--Klein
scenario this can be done merely by adding a small mass term to the
higher-dimensional action. As we will see shortly, this apparently
innocent step has non-trivial consequences in the domain wall
case. The point is that even in the presence of an explicit bulk mass
term, the operator which determines the modes (and corresponding
eigenvalues, i.e., masses of 4-dimensional particles) always has a
continuous spectrum starting from zero. Indeed, consider the case of a
free 5-dimensional massive scalar field described by the action
\begin{equation}
S=\int dzd^4x\sqrt{-g}\left(
{1\over 2}g^{ab}\partial_a\phi\partial_b\phi
-{1\over 2}\mu^2\phi^2\right)\;, 
\label{scal-action}
\end{equation}
where the metric $g^{ab}$ is given by eq. (\ref{RSmetric}). The field
equation in this background reads 
\begin{equation}
\label{scalar}
\left[-\partial_z^2+4k\,{\rm sign}(z)\partial_z
+\mu^2-m^2\e^{2k|z|}\right]\phi(z,p) =0\;,
\end{equation}
where $m^2=p^{\mu}p_{\mu}$ is the four-dimensional mass. Clearly, at
large $|z|$ the bulk mass term $\mu^2$ is negligible as compared to
the term $m^2\e^{2k|z|}$, so equation (\ref{scalar}) reduces to one
with $\mu=0$. Since the continuum eigenvalues are determined by the
large-$|z|$ asymptotics which is not affected by the bulk mass term
$\mu^2$, equation (\ref{scalar}) has the same continuum spectrum as
the equation with $\mu=0$, i.e., continuum spectrum starts from zero
$m$.

Obviously, there is no zero mode at $\mu \neq 0$. The above argument
shows that there are no true localized modes with non-zero
4-dimensional mass $m$ either (there are no true bound states embedded
in the continuum). This property is generic to fields that can have
bulk modes: by scaling argument, bulk mass terms are suppressed by a
factor $a^2(z)$ as compared to $p^2$, hence they become irrelevant at
large distances from the wall.  We will see this explicitly in the
case of fermions as well.  Let us note in passing that the phenomenon
we are discussing persists also when the mass terms are introduced on
the brane itself (say, when the effective action contains an
additional term $\int~d^4x dz \delta(z)[-(1/2)\sqrt{(-\bar{g})}
\mu^{\prime 2} \phi^2]$ due to some dynamics on the brane).

A question arises whether the domain wall scenario is at all capable
of incorporating objects which, to a certain approximation, behave
like 4-dimensional particles of small, but non-vanishing mass.  In
this paper we give an affirmative answer in both scalar and fermion
cases. We will see, however, that these 4-dimensional particles are
metastable. In other words, we show that at small enough $\mu$, there
exist quasi-localized modes whose width $\Gamma$ is much smaller than
their 4-dimensional mass $m$.  These quasi-localized modes are
metastable states that decay into the continuum modes. From the point
of view of 4-dimensional observer, the quasi-localized modes
correspond to massive particles that propagate in three spatial
dimensions for some time, and then literally disappear (into the fifth
dimension).

Quasi-localized scalars and fermions are similar to quasi-localized
gravitons \cite{GRS,Csaki1,Dvali1} that emerge in a class of models
\cite{CGR,GRS} with flat large-$z$ asymptotics of the 5-dimensional
space-time. Unlike the latter, the models we consider need not contain
potentially dangerous \cite{Dvali1,Witten,Dvali2,Ratazzi} dynamical
branes of negative tension.

The suppression of the width $\Gamma$ depends on the mechanism of the
localization of particles on the wall.  We find that in the scalar
model (\ref{scal-action}), the width is suppressed with respect to the
mass $m$ by a factor $(m/k)^2$ at small $m/k$. In the case of fermions
the suppression factor has more complicated form and is exponential in
a range of parameters.

Yet another manifestation of the continuum starting from zero $m$ is a
power law behaviour of the 4-dimensional propagator in the
infrared. In the scalar case this corresponds to a power-law potential
between static sources at large distances (in a model meant to
describe massive 4-dimensional particles!). We will explicitly
calculate this potential in section 2.

\section{Scalar field}

There are several ways to see that an effective 4-dimensional theory
contains a massive metastable particle. The easiest way is to directly
find a complex eigenvalue from the equation which determines the mass
spectrum (eq.(\ref{scalar}) in the case of scalars). As in ordinary
quantum mechanics, this complex eigenvalue appears when one imposes
the radiation (outgoing wave) boundary conditions\footnote{In the
brane-world context, this approach was used in ref.\cite{Dvali1} for
calculating the lifetime of quasi-localized gravitons in models of the
type of Refs.\cite{CGR,GRS}} at $z\to \pm \infty$. Alternatively, one
calculates the Feynman propagator between two points on the brane: if
there exists a metastable state, this propagator has a pole at a
complex value of the mass.  The two ways should of course lead to
consistent results.

Let us begin with applying the first method to scalars.  We wish to
show that the mode equation, eq.(\ref{scalar}), has a complex
eigenvalue when the radiation boundary conditions are imposed
(cf. ref. \cite{GKR}) at $z\to \pm \infty$.  To the left and to the
right of the brane, the solutions to eq.(\ref{scalar}) which satisfy
the radiation boundary conditions are:
\begin{eqnarray}
f(z<0) &=& c_1 \e^{-2kz} H^{(1)}_{\nu}\left({m\over k}\e^{-kz}\right),
\nonumber\\
f(z>0) &=& c_2 \e^{2kz} H^{(1)}_{\nu}\left({m\over k}\e^{kz}\right),
\label{solutionssc}
\end{eqnarray}
where $H^{(1)}_{\nu}(x)$ is the Hankel function and 
\[
\nu = \sqrt{4+{\mu^2\over k^2}}. 
\]
The eigenvalues are determined by matching these solutions at
$z=0$. The continuity requires that $c_1=c_2$. The first derivative
should also be continuous, as is clear from eq.(\ref{scalar}),
\[
\partial_z f(+0) - \partial_z f(-0) = 0.
\]  
The latter condition implies the equation for the eigenvalue $m$,
\begin{equation}
{mH_{\nu-1}^{(1)}( {m/k})\over k H_{\nu}^{(1)}( {m/k})} + 2 -\nu = 0.
\label{pole-eq}
\end{equation}
Let us consider the case $\mu\ll k$, and search for solutions with
$m\ll k$. In this regime one writes
\[
{H_{\nu-1}^{(1)}(m/k)\over H_{\nu}^{(1)}(m/k)}= 
{N_{\nu-1}(m/k)\over N_{\nu}(m/k)} \left\{ 
1-i{J_{\nu-1}(m/k)\over N_{\nu-1}(m/k)} + \ldots \right\},
\]
where dots denote terms suppressed by at least one power of $m/k$, and
we keep the contribution that is imaginary at real $m$. Plugging this
expression into eq.(\ref{pole-eq}) and expanding the Bessel functions
at small argument one finds
\[
m=m_0-i\Gamma
\]
with 
\begin{eqnarray}
m_0^2 &=& {\mu^2\over 2} \quad,\label{mass}\\
\frac{\Gamma}{m_0}&=& {\pi\over 16}\left( {m_0\over k}\right)^2 \quad.
\label{width}
\end{eqnarray}
Thus, there exists a quasi-discrete level with the width suppressed by
$(m/k)^2$. 

It is instructive to reproduce this result in terms of the scalar
propagator $\Delta(z,z',p^2)$ (by 4-dimensional Lorentz invariance,
the latter depends only on $p^2=p^{\mu}p_{\mu}$). The pole of the
propagator at certain $p^2=m^2$ corresponds to a particle with the
4-dimensional mass $m$. In the case when one of the arguments is
located on the brane, the propagator is straightforward to find from
eq.(\ref{solutionssc}),
\[
\Delta(z,0,p^2)=c(p) \e^{2k|z|}H_{\nu}^{(1)}\left( {p\over k}
\e^{k|z|}\right)\;,
\]
where the radiation boundary conditions are imposed, and $c(p)$ is
determined by the normalization condition
\[
\partial_z \Delta(z,0,p^2)\Bigm|_{z=0}=1\;.
\]
The propagator has a particularly simple form when both arguments are
on the brane,
\begin{equation}
\label{propagator}
\Delta(0,0,p^2)=\left[ {pH_{\nu-1}^{(1)}(p/k)\over
k H_{\nu}^{(1)}(p/k)}+2-\nu \right]^{-1} \;.
\end{equation}
Comparing eq.(\ref{propagator}) with eq.(\ref{pole-eq}) one finds that
the propagator has the pole at the complex value of $p^2$ which
corresponds to the unstable massive particle with the mass and width
given by eqs.(\ref{mass}) and (\ref{width}).

Finally, let us consider the static potential between two sources on
the brane, which is induced by the scalar exchange. The potential
recieves contributions from all modes and is given by the following
integral
\begin{equation}
\label{potential}
V(r)=q_1q_2\int{\e^{-mr}\over r}\phi_m^2(0)dm\;,
\end{equation}
where $q_1$ and $q_2$ are the charges of the sources and $\phi_m(z)$
are the eigenmodes of eq.(\ref{scalar}) which are even under the
reflection $z\to -z$. These eigenmodes are normalized with the measure
$\exp(-2k|z|)$ \cite{Giga,GKR},
\[
\int dz \e^{-2k|z|} \phi_m(z)\phi_{m'}(z)= \delta(m-m').
\]
One finds 
\begin{equation}
\label{mode}
\phi_m(z)=\sqrt{{\pi m\over 2k}}\e^{2k|z|}
\left[a_m J_{\nu}\left({m\over k}\e^{k|z|}\right)+
b_m N_{\nu}\left({m\over k}\e^{k|z|}\right)\right]\;,
\end{equation}
where the coefficients $a_m$ and $b_m$ are determined by the
normalization condition
\[
a_m^2+b_m^2=1
\]
and  the boundary condition on the brane,
\[
\partial_z \phi_m(z)\Bigm|_{z=0}=0\;.
\]
The solution to these equations can be written in the form
\[
a_m=-{A_m\over \sqrt{1+A_m^2}} \,\,\, , \, \quad 
b_m={1\over \sqrt{1+A_m^2}}
\]
where
\[
A_m={N_{\nu-1}\left({m\over k}\right)
-(\nu-2){k\over m}N_{\nu}\left({m\over k}\right)
\over J_{\nu-1}\left({m\over k}\right)
-(\nu-2){k\over m}J_{\nu}\left({m\over k}\right)}\;.
\]
At relatively large distances, $r\gg k^{-1}$, only modes with $m\ll k$
are relevant. Assuming again that $\mu\ll k$, we find
\[
A_m\approx {2\Gamma(\nu+1)\Gamma(\nu-1)\over\pi(\nu+2)}\left({m\over
2k}\right)^{2-2\nu} 
\left[ 1 - 2(\nu-2)(\nu-1)\left({k\over
m}\right)^2\right] \;,
\]
\[
\phi_m^2(0) \approx {1\over2\pi}\left( m\over
k\right)^{-3}{16\over 1+A_m^2},
\]
so the scalar potential (\ref{potential}) takes the following form
\begin{equation}
\label{potential1}
V(r)={8q_1q_2\over\pi}\int{\e^{-mr}\over r}\left({m\over
k}\right)^{-3}{1\over 1+A_m^2}dm\;.
\label{intpot}
\end{equation}
There are two competing contributions to this integral. The first one
comes from the region where $A_m$ are small, i.e., the last factor in
the integrand of eq.(\ref{intpot}) is peaked.  It is straightforward
to check that this region corresponds exactly to the resonance
(\ref{mass}), (\ref{width}) described above. The resonance
contribution to the potential is equal to
\begin{equation}
\label{respot}
V_{\rm res}(r) = {8q_1q_2\over\pi}{\e^{-m_0r}\over r}\left({m_0\over
k}\right)^{-3}2\pi\Gamma=\pi q_1q_2k{\e^{-m_0r}\over r}\;.
\end{equation}
As one might have expected, this is the usual Yukawa potential with
the mass $m_0$ (extra factor $k$ accounts for the difference in the
mass dimensions of charges in five and four dimensions).

The second contribution comes from the light modes with $m\ll \mu$. 
It is suppressed by the large factor $(1+A_m^2)\propto (k\mu)^4/m^8$.
Explicitly, 
\begin{equation}
V_{\rm light}(r)={\pi q_1q_2\over 2}\int{\e^{-mr}\over r}{m^5\over
km_0^4}dm=60\pi q_1q_2 \cdot {1\over km_0^4} \cdot \frac{1}{r^7}\;.
\label{light}
\end{equation}
We see that almost massless modes lead to power-law behavior at large
$r$. The resulting potential
\[
V(r) = V_{\rm res}(r) + V_{\rm light}(r)
\]
is dominated by the power-like contribution at distances 
$r\gtrsim 2m_0^{-1}\ln(k/m_0)$.

\section{Fermions}

Fermion fields are not localized on the positive tension brane by
gravitational interactions only \cite{Giga}.  Hence, one invokes the
localization mechanism of Refs.\cite{JackiwRebbi,RS}. The simplest
set up is as follows. One considers a domain wall formed by some
scalar field $\chi$. This scalar field has a double-well potential
with two degenerate vacua at $\chi=\pm v$; the domain wall separates
the region $\chi=-v$ at $z<0$ from the region $\chi=v$ at $z>0$.  A
fermion field which has a Yukawa coupling to the scalar,
$g\chi\bar\psi\psi$, has an exact zero mode in the domain wall
background. This zero mode is topological and its existence does not
depend on the details of the profile of the scalar filed across the
wall. Therefore, it also exists for the infinitely thin wall, which is
the case we consider in what follows.

For a given sign of the Yukawa coupling $g$, the zero mode has a
certain chirality. Since the 4-dimensional fermion mass term requires
both chiralities, it can only be introduced in models with two bulk
fermion fields which have opposite couplings to the scalar $\chi$.  It
is convenient to organize these spinors into one field
\[
\Psi = \left(
\begin{array}{c}
\psi_1\\\psi_2
\end{array} \right).
\]
where $\psi_1$ and $\psi_2$ are four-component spinors living in 
five dimensions.

In the presence of the Yukawa interaction, $g\chi\bar\Psi\tau_3\Psi$,
the fields $\psi_1$ and $\psi_2$ have left and right zero modes,
respectively. Mixing between these two modes that eventually gives
rise to 4-dimensional mass, is introduced by adding a term
$\mu\bar{\Psi}\tau_1\Psi$. It is convenient to bring both these terms
to the off-diagonal form by a global $SU(2)$ rotation. The resulting
fermion action reads
\[
S = \int dz\,d^4x\sqrt{g}\, \bar\Psi \left( i  \gamma^a\nabla_a 
+  g\chi \tau_1 + \mu\tau_2\right)\Psi, 
\]
where $\nabla_a$ is the spinor covariant derivative with respect to
the 5-dimensional metric $g_{ab}$. The Dirac equation which follows
from this action in the background (\ref{RSmetric}) has the form
\begin{equation}
\left[ {1\over a} \gamma^{\mu}p_{\mu} 
+ \gamma_5 \partial_z - 
g\chi(z) \tau_1 - \mu\tau_2\right] \Psi = 0.
\label{dirac}
\end{equation}
In the thin-wall limit one has 
    $g\chi(z) = gv\, {\rm sign} z$.

Equation (\ref{dirac}) determines the fermion modes.  At $\mu=0$ and
$gv > k/2$ \cite{Giga}, there exist two fermion zero modes of opposite
chirality and continuous spectrum starting from zero. It is
straightfroward to see that at $\mu>0$ the zero modes disappear,
whereas the continuous spectrum still starts from zero. This is
precisely the same situation as in the scalar case.

In order to see that there is a metastable massive state, let us find
the complex eigenvalue at which there exists a solution to
eq.(\ref{dirac}) with the radiation boundary conditions imposed at $z
\to \pm \infty$.  It is convenient to separate the spinor $\Psi$ into
the left and right components,
\[
\gamma_5\Psi_{L,R} = \pm \Psi_{L,R}.
\]
In terms of $\Psi_{L,R}$ eq.(\ref{dirac}) translates into a set of
coupled equations,  
\begin{eqnarray}
\label{eq-psiR}
{(\gamma p)\over a}  \Psi_R + \partial_z \Psi_L 
- (g\chi\tau_1 + \mu\tau_2) \Psi_L &=& 0,\\
\label{eq-psiL}
{(\gamma p)\over a}  \Psi_L - \partial_z \Psi_R 
- (g\chi\tau_1 + \mu\tau_2) \Psi_R &=& 0.
\end{eqnarray}
After eliminating $\Psi_R$ one obtains a 
second order equation for $\Psi_L$, 
\begin{equation}
\left[
{m^2\over a^2} + \partial_z^2 + {a'\over a} \partial_z 
- {a'\over a} (g\chi\tau_1 + \mu\tau_2) - g \chi' \tau_1- 
 (g^2\chi^2 + \mu^2)\right] \Psi_L = 0.
\label{eq-secondorder}
\end{equation}
Again, the explicit mass terms (the terms involving $g\chi$ and $\mu$)
are negligible at large $|z|$ as compared to $m^2/a^2$, and continuum
indeed starts at zero $m$.

We solve eq.(\ref{eq-secondorder}) separately to the left and to the
right of the brane, and then match the solutions.  To the right of the
brane one has $a=\exp(-kz)$, so eq. (\ref{eq-secondorder}) reads
\begin{equation}
\left[
m^2e^{2kz} + \partial_z^2 -k \partial_z 
+k (gv\tau_1 + \mu \tau_2) - 
 (g^2v^2 + \mu^2) \right] \Psi_L = 0
\label{10*}
\end{equation}
It is convenient to introduce eigenvectors of the matrix $(gv\tau_1 +
\mu \tau_2)$. Let us define $M$ and $\alpha$ in such a way that
\[
gv +i\mu = M\e^{i\alpha}. 
\]
Then these eigenvectors are
\[
\Psi_{\pm}^{(>)} = \left(
\begin{array}{c}
e^{-i\alpha/2}\\ \pm e^{i\alpha/2}
\end{array}\right)
\]
with the eigenvalues $\pm M$. 

It is now straightforward to obtain a general solution to
eq.(\ref{10*}) that obeys the radiation boundary conditions at $z \to
+ \infty$,
\[
\Psi_L(z>0) = \e^{\frac{kz}{2}} 
\left[ 
c^{(>)} H^{(1)}_{\nu_{+}} \left(\frac{m}{k}\e^{kz}\right)\Psi_{+}^{(>)} +
d^{(>)} H^{(1)}_{\nu_{-}} \left(\frac{m}{k}\e^{kz}\right) \Psi_{-}^{(>)}
\right]\psi_p\;,
\]
where
\[
\nu_{\pm} = {M\over k} \mp {1\over 2}\;,
\]
$c^{(>)}$ and $d^{(>)}$ are two yet undetermined coefficients and
$\psi_p$ is a $z$-independent left spinor. 

The solution to the left of the brane is obtained in a similar
way,
\[
   \Psi_L(z<0) = \e^{-\frac{kz}{2}} 
\left[ 
c^{(<)} H^{(1)}_{\nu_{+}} \left(\frac{m}{k}\e^{-kz}\right)\Psi_{+}^{(<)} +
d^{(<)} H^{(1)}_{\nu_{-}} \left(\frac{m}{k}\e^{-kz}\right) \Psi_{-}^{(<)}
\right]\psi_p\;,
\]
where
\[
\Psi_{\pm}^{(<)} = \left(
\begin{array}{c}
e^{i\alpha/2}\\ \pm e^{-i\alpha/2}
\end{array}\right)
\]
are eigenvectors of the matrix $(-gv\tau_1 + \tau_2)$.

The fermion wave function has to obey matching conditions at $z=0$.
These are the requirements of continuity of $\Psi_L$ and $\Psi_R$
across the brane,
\begin{equation}
\Psi_{L,R}(-0) = \Psi_{L,R}(+0). 
\label{matching-psi}
\end{equation}
Continuity of the left components requires
\begin{eqnarray}
\gamma c^{(>)}  + d^{(>)} &=& \exp(i\alpha)(\gamma c^{(<)} + d^{(<)}),
\label{mat1}
\\
\gamma c^{(>)} - d^{(>)} &=& \exp(-i\alpha)(\gamma c^{(<)} - d^{(<)}),
\label{mat2}
\end{eqnarray}
where we have introduced the notation
\[
\gamma \equiv {H_{\nu_{+}}(m/k)\over H_{\nu_{-}}(m/k)}.
\]
To obtain the second set of relations between $c$'s and $d$'s, one
notices that, because of eq.(\ref{eq-psiR}), continuity of $\Psi_R$
across the brane is equivalent to continuity of
\[
\partial_z \Psi_L 
- (g\chi\tau_1 + \mu\tau_2) \Psi_L 
\]
Making use of the properties of Hankel functions, we obtain
\begin{eqnarray}
c^{(>)}  - \gamma d^{(>)} &=& \exp(i\alpha)(- c^{(<)} + \gamma d^{(<)}),
\label{mat3}
\\
c^{(>)}+ \gamma d^{(>)} &=& \exp(-i\alpha)(- c^{(<)} - \gamma d^{(<)}).
\label{mat4}
\end{eqnarray}
The determinant of the system (\ref{mat1}) -- (\ref{mat4})
vanishes provided that $\gamma$ obeys either of the four equations,
\[
\gamma = \pm {\rm tan}(\alpha/2),\qquad 
\gamma = \pm {\rm cot}(\alpha/2) .
\]
At small $\mu / gv$ (i.e., small $\alpha$), the relevant solution is
$\gamma = {\rm tan} (\alpha/2)$. This equation determines the complex
eigenvalue $m$. Explicitly, the eigenvalue equation at $\mu \ll gv$
reads
\[
{H^{(1)}_{\nu_+}(m/k)\over H^{(1)}_{\nu_-}(m/k)}
= \frac{\mu}{2gv}
\]
The simplest case to consider is when $\mu$ is the smallest parameter,
i.e., $\mu \ll k$.
In this case one expands the Bessel functions at small values
of the argument and obtains
\[
m=m_0-i\Gamma
\]
with 
\[
m_0 = \left(1 -{k\over 2gv}\right) \mu 
\]
\[
{\Gamma\over m_0} = \left( {m_0\over 2k}\right)^{2gv/k-1}
{\pi \over [\Gamma(gv/k+1/2)]^2}
\]
Hence, the suppression of the width depends non-trivially on all
parameters and may become very strong. 

In the opposite case $\mu \gg k$ (but still $\mu \ll gv$, which
implies also $gv \gg k$), one makes use of the approximation of the
Bessel function by means of tangents, and obtains
\[
m_0 = \mu \, ,
\]
\[
\frac{\Gamma}{m_0} = \frac{1}{2} 
\left(\frac{m_0}{2M}\right)^{2M/k -1} \e^{2M/k}
\]
where $M = \sqrt{(gv)^2 + \mu^2}$.
In this case the suppression of the width is always exponentially
strong. One can show that at $\mu \gg k$, the width is exponentially 
suppressed also for $\mu \sim gv$,
\[
\frac{\Gamma}{m_0} \propto 
\e^{-2\frac{M}{k}(\beta - {\rm tanh}\beta)}\,\,\, , \, \quad
     {\rm cosh} \beta = \frac{M}{\mu} = \frac{M}{m_0} \quad .
\]
It is clear why the time fermions spend on the brane is large at small
$k$. At $gv \gg k$, continuum modes with $p^2 \sim m_0^2$ barely
penetrate the potential barrier extending from the brane to the
large-$z$ region. The would-be localized mode, on the other hand, is
narrow in $z$ direction ($\Delta z \sim (gv)^{-1}$).  Hence, the
overlap between continuum modes and would-be localized mode is small,
and the lifetime of the metastable state is large.  This feature is
absent in the scalar case considered in section 2, where both the
potential barrier and the spatial extent of the would-be localized
mode are governed by one and the same parameter $k$.

Peculiar features of massive matter in brane world have been found in
this paper in field theory framework. It remains to be understood
whether similar phenomena are present in D-brane theory. In particular
one may wonder whether massive matter carrying gauge charges may
disappear into extra dimensions. One may worry that this would
contradict 3-dimensional Gauss' law; however, the issue becomes not so
obvious if one recalls that the gravitational analog of Gauss' law
does not prevent massive particles to escape from the brane
\cite{GRS2}. We hope to return to this and other related issues 
in future.

We are indebted to F.~Bezrukov, W.~Buchm\"uller, T.~Gherghetta,
M.~Li\-ba\-nov, M.~Shaposhnikov and S.~Sibiryakov for helpful
discussions. S.D. thanks Institute of Theoretical Physics, University
of Lausanne, where part of this work has been done, for hospitality.
This work is supported in part by RFBR grant 99-02-18410 and by the CRDF
grant 6603. The work of S.D. is also supported by the Russian Academy
of Sciences, JRP grant 37 and by ISSEP fellowship. The work of P.T. is
supported in part by the Swiss Science Foundation, grant 21-58947.99.

\end{document}